\documentclass[aps,prb,twocolumn]{revtex4}
\usepackage{graphicx}
\begin{document}

\title{\emph{Ab initio} potential energy surfaces, bound states
and electronic spectrum of the Ar--SH complex}
\author{Richard J. Doyle}
\affiliation{Department of Chemistry, University of Sheffield,
Sheffield S3 7HF, United Kingdom}
\author{David M. Hirst}
\affiliation{Department of Chemistry, University of Warwick,
Coventry CV4 7AL, United Kingdom}
\author{Jeremy M. Hutson}
\affiliation{Department of Chemistry, University of Durham, South
Road, Durham DH1 3LE, United Kingdom}

\date{\today}

\begin{abstract}
New \emph{ab initio} potential energy surfaces for the $^2\Pi$
ground electronic state of the Ar--SH complex are presented,
calculated at the RCCSD(T)/aug-cc-pV5Z level. Weakly bound
rotation-vibration levels are calculated using coupled-channel
methods that properly account for the coupling between the two
electronic states. The resulting wavefunctions are analysed and a
new adiabatic approximation including spin-orbit coupling is
proposed. The ground-state wavefunctions are combined with those
obtained for the excited $^2\Sigma^+$ state [Phys. Chem. Chem. Phys.
6, 5463 (2004)] to produce transition dipole moments. Modelling the
transition intensities as a combination of these dipole moments and
calculated lifetime values [J. Chem. Phys. 109, 170 (1998)] leads to
a good representation of the experimental fluorescence excitation
spectrum [J. Chem. Phys. 98, 4301 (1993)].

\end{abstract}
\maketitle

\section{Introduction}

Van der Waals complexes containing open-shell species are of great
current interest. In particular, complexes containing atoms or
molecules with orbital angular momentum necessarily involve
multiple electronic states.\cite{DUBERNET:1991,
DUBERNET:open:1994} They provide a test-bed for studying
electronically non-adiabatic effects, which are important in the
theory of reaction dynamics. \cite{alexander2004, retail2005,
Ziemkiewicz:2005} In addition, the observation of pre-reactive
Van der Waals complexes trapped in bound levels \cite{Loomis:1997,
Liu:1999, wheeler2000, Lester:2001, Merritt:2005} can shed light
on intermolecular forces in the entrance and exit channels of
chemical reactions.\cite{Dubernet:clhcl:1994, Meuwly:2003,
klos2004, Fishchuk:2006} The form of these shallow, long-range
wells can be important in determining reaction outcomes
\cite{Skouteris:1999, wang2002} and transition-state geometries.
\cite{neumark2002, Neumark:2005}

In this paper, we consider the complex consisting of an open-shell
SH radical and an Ar atom. New \emph{ab initio} potential energy
surfaces (PESs) for the \emph{X} $^2\Pi$ state are presented, and
used in calculations of the bound rotation-vibration levels. We
discuss the possibility of employing an approximate approach in
the bound-state calculations, using a single adiabatic PES rather
than the two surfaces used in the standard method.
\cite{DUBERNET:1991} Finally, the bound-state energies and
wavefunctions are used to simulate the vibrationally resolved
electronic spectrum.

The Ar--SH cluster was first detected experimentally by Miller and
co-workers \cite{miller92} using laser-induced fluorescence
excitation spectroscopy. Subsequently this group developed
empirical potential energy surfaces for the complex, for both the
\emph{A} state \cite{miller97} and the \emph{X} state,
\cite{miller99} by fitting model functions to reproduce
laser-induced fluorescence (LIF) results. The region of the
\emph{A} state PES corresponding to the Ar--S-H configuration
(Jacobi angles between $\sim 90$ and 180 degrees) was determined
only approximately, because the fluorescence experiments did not
probe this zone. More recently, Hirst \emph{et al.}\
\cite{hirst2004} have presented a PES for the \emph{A} state based
on \emph{ab-initio} calculations at the RCCSD(T) level with the
aug-cc-pV5Z basis set. This surface was used to predict bound
vibrational levels in the Ar--S-H configuration \cite{hirst2004}
which have not, to our knowledge, been observed in experiment so
far. A possible reason why these levels have eluded detection is
discussed in section IV of this paper.

Sumiyoshi \emph{et al.}\ \cite{sumiyoshi2000} have recorded
high-resolution spectra for Ar--SH in the ground electronic state
using Fourier-transform microwave (FTMW) spectroscopy. These
authors also produced PESs for the \emph{X} state based on fitting
a function to reproduce their experimental results,
\cite{sumiyoshi2000} and these surfaces were later improved with
the aid of some \emph{ab initio} results.\cite{sumiyoshi2003}
Most recently, results from microwave-millimeter-wave
double-resonance spectroscopy \cite{sumiyoshi2005i} were employed
to determine new three-dimensional PESs for the \emph{X} state.
\cite{sumiyoshi2005ii}

The family of weakly bound clusters containing a rare gas atom and
either the OH or SH radical have been reviewed by Carter \emph{et
al.} \cite{miller2000} in 2000 and by Heaven \cite{heaven2005} in
2005.

The structure of the present paper is as follows. In Section II we
present new PESs for Ar--SH ($^2\Pi$) based entirely on \emph{ab
initio} calculations at the RCCSD(T) level with an aug-cc-pV5Z
basis set. In Section III we describe bound rotation-vibration
level calculations using these surfaces. We also investigate the
wavefunctions and introduce a new adiabatic approximation for the
bound states, including spin-orbit coupling. In Section IV the
results are combined with those of a previous study of the
\emph{A} $^2\Sigma^+$ state, in order to produce a high-quality
simulation of the vibrationally resolved electronic spectrum.

\section{Potential energy surfaces}
The geometry of the complex is specified in terms of body-fixed
Jacobi coordinates $r$, $R$ and $\theta$. $R$ is the length of the
vector $\mathbf{R}$ which links the center of mass of the SH
fragment to the Ar nucleus. The vector $\mathbf{r}$ links the S
nucleus to the H nucleus: its modulus $r$ is the SH bond length.
The angle between $\mathbf{R}$ and $\mathbf{r}$ is $\theta$, so
that $\theta=0$ corresponds to a linear Ar--H-S configuration. For
this work the bond length $r$ was held constant at the
experimentally determined equilibrium value of 1.3409 \AA,\
\cite{herzbergIV} which is justified because the vibrational
motion of the diatom is very weakly coupled to the relatively
low-frequency Van der Waals modes of interest.

Energies were calculated for a regular grid of geometry points
using the MOLPRO quantum chemistry program.\cite{molpro} These
points are at every distance $R$ from 3.25 {\AA } to 5.5 {\AA } in
steps of 0.25 {\AA } and for every angle $\theta$ from 0$^{\circ}$
to 180$^{\circ}$ in steps of 15$^{\circ}$. This gives a total of
130 points. We used the RCCSD(T) method (restricted coupled
cluster with single, double and non-iterative triple excitations)
\cite{knowles93, knowles2000} with the aug-cc-pV5Z basis set.
\cite{dunning89, dunning92, dunning94} The counterpoise procedure
of Boys and Bernardi \cite{boys70} was used to correct for basis
set superposition error (BSSE). This is the same level of theory
and basis set as were used in recent calculations of the PES for
the $A$-state of Ar--SH.\cite{hirst2004}

\begin{figure}
\includegraphics[width=85mm]{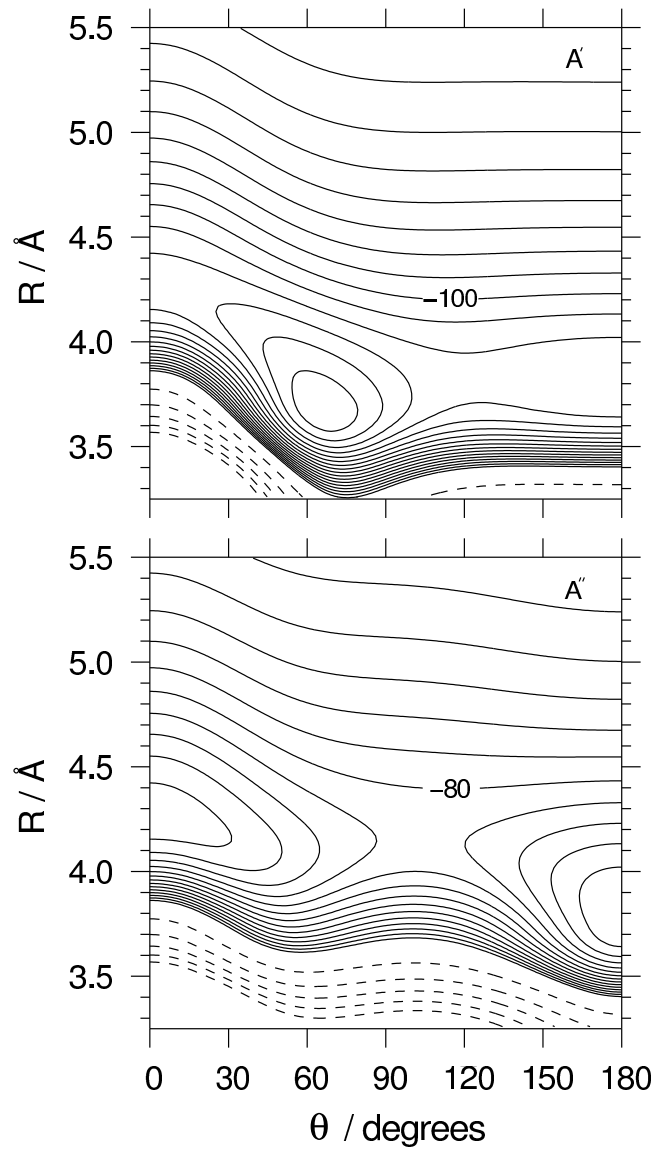}
\caption{\emph{Ab initio} potential energy surface contour plots
for Ar--SH ({\em X}) in the $A'$ state (upper plot) and the $A''$
state (lower plot). Solid contour lines are shown at 10 cm$^{-1}$
intervals, ranging from 0 cm$^{-1}$ to $-150$ cm$^{-1}$ inclusive
for the $A'$ surface and 0 cm$^{-1}$ to $-120$ cm$^{-1}$ inclusive
for the $A''$ surface. Dashed contour lines are shown at 100
cm$^{-1}$ intervals from +100 cm$^{-1}$ to +500 cm$^{-1}$
inclusive for both surfaces. The linear Ar--H-S conformation
corresponds to $\theta $ = 0$^{\circ}$. \label{fig:adiabaticPESs}}
\end{figure}

Two potential surfaces were obtained from the \emph{ab initio}
calculations. These correspond to two adiabatic electronic states:
one symmetric ($A'$) and one antisymmetric ($A''$) with respect to
reflection in the plane of the nuclei. The two states are
degenerate at linear geometries but nondegenerate at nonlinear
geometries: the splitting is an example of the Renner-Teller
effect. The interaction energies for each state were interpolated
using a 2D spline function, and contour plots of the resulting
surfaces are shown in Fig.\ \ref{fig:adiabaticPESs}. The $A'$ and
$A''$ surfaces result from the electronic Hamiltonian without
spin-orbit coupling: a discussion of surfaces including spin-orbit
coupling is presented in section IV.

The adiabatic surfaces (adiabats) are qualitatively similar to those
reported for He--SH and Ne--SH complexes.\cite{Cybulski:2000} The
latter were calculated at the RCCSD(T) level, using the smaller
aug-cc-pVTZ basis set, but with an additional set of bond functions,
and counterpoise correction. A comparison of the positions and
energies of the minima on the Ar--SH surfaces presented here with
those for $X$-state He--SH and Ne--SH is given in Table \ref{tab1}.
For all the $A''$ surfaces there is a global minimum in the linear
X--S-H configuration (where X is He, Ne or Ar) and a local minimum
in the linear X--H-S configuration. The X--H-S configurations are
saddle points on the $A'$ surfaces, which have shallow local minima
at $\theta=180^\circ$ and global minima at nonlinear configurations.
The $A'$ global minima are deeper than those on the $A''$ surfaces,
because in the $A'$ state the SH $\pi$ hole is directed towards the
Ar atom, resulting in reduced repulsion. The global minima for the
$A'$ state occur at angles $\theta$ that increase with the atomic
number of the rare gas atom. Also, as expected, the minima are
deeper for clusters containing heavier (and more polarizable) rare
gas atoms.

\begin{table}
\caption{\label{tab1} Positions and well depths of potential
minima on the $A'$ and $A''$ adiabatic surfaces for the $X$-state
of SH--rare gas clusters. The results for Ne--SH and He--SH
clusters are from ref.\ \onlinecite{Cybulski:2000}.}
\begin{tabular}{ccccc}
\hline\hline
Cluster & State & $R$/\AA & \qquad $\theta/^{\circ}$ \qquad & depth/cm$^{-1}$\\
\hline
Ar--SH & $A'$ & 3.678 & 66.6 & 157.69\\
Ne--SH & $A'$ & 3.611 & 57.2 & 57.05\\
He--SH & $A'$ & 3.639 & 54.4 & 25.97\\
\hline
Ar--SH & \quad $A'$ and $A''$ \quad & 3.801 & 180 & 128.54\\
Ne--SH & \quad $A'$ and $A''$ \quad & 3.593 & 180 & 54.27\\
He--SH & \quad $A'$ and $A''$ \quad & 3.593 & 180 & 25.27\\
\hline
Ar--SH & $A''$ & 4.274 & 0 & 125.22\\
Ne--SH & $A''$ & 4.101 & 0 & 45.75\\
He--SH & $A''$ & 4.126 & 0 & 21.16\\
\hline\hline
\end{tabular}
\end{table}

\begin{figure}
\includegraphics[width=85mm]{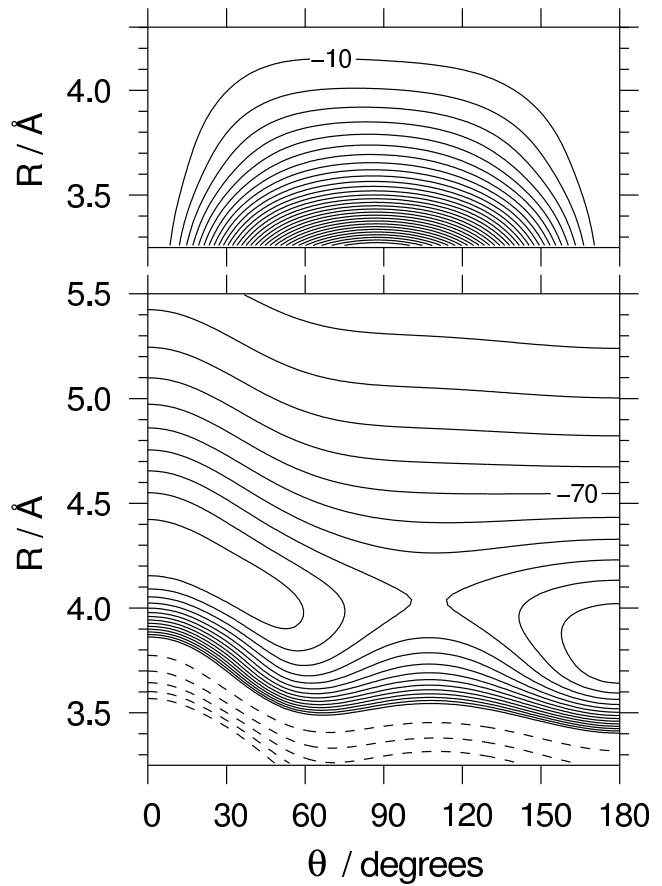}
\caption{Contour plots for the Ar--SH ({\em X}) difference (upper
plot) and sum (lower plot) potential energy surfaces. Solid
contour lines are shown at 10 cm$^{-1}$ intervals, ranging from 0
cm$^{-1}$ to $-300$ cm$^{-1}$ inclusive for the difference surface
and from 0 cm$^{-1}$ to $-120$ cm$^{-1}$ inclusive for the sum
surface. Dashed contour lines are shown at 100 cm$^{-1}$ intervals
from +100 cm$^{-1}$ to +500 cm$^{-1}$ for the sum surface only.
The linear Ar--H-S conformation corresponds to $\alpha$ =
0$^{\circ}$. \label{fig:avdiffpots}}
\end{figure}

In order to perform dynamical calculations on Ar--SH, we need to
evaluate the matrix elements of the potential between electronic
states labelled with an angular momentum quantum number $\lambda$.
For this purpose it is convenient to re-express the potential
energy surfaces as the sum ($V_0$) and difference ($V_2$)
potentials
\begin{displaymath}
V_0(R,\theta)=\frac{1}{2}\left[{V_{A'}(R,\theta)+V_{A''}(R,\theta)}\right],
\end{displaymath}
\begin{displaymath}
V_2(R,\theta)=\frac{1}{2}\left[{V_{A'}(R,\theta)-V_{A''}(R,\theta)}\right].
\end{displaymath}
Contour plots of these surfaces are shown in Fig.\
\ref{fig:avdiffpots}. They are quite similar to those recently
presented by Sumiyoshi \emph{et al.}\ \cite{sumiyoshi2005ii} The
latter were fitted to reproduce experimental results, with
starting values for the potential parameters obtained from
\emph{ab initio} calculations (RCCSD(T)/aug-cc-pVTZ). The form of
our sum potential is also qualitatively similar to those recently presented
for Ne--SH and Kr--SH by Suma \emph{et al.}\cite{suma04}

\section{Bound-state calculations}

\subsection{Coupled channel calculations}

The bound states of a complex such as Ar--SH (\emph{X}$^2\Pi$)
involve \emph{both} potential energy surfaces. In the present work
we use a coupled-channel approach to calculate the bound states.
In a body-fixed axis system the Hamiltonian operator is
\begin{equation}
H= -\frac{\hbar^2}{2\mu}R^{-1}\left(\frac{\partial^2}{\partial
R^2}\right) R + H_{\rm mon} + \frac{\hbar^2(\hat J
-\hat\jmath)^2}{2\mu R^2}+\hat V, \label{ham}
\end{equation}
where $H_{\rm mon}$ is the monomer Hamiltonian and $\hat V$ is the
intermolecular potential. In a full treatment including overall
rotation, the total wavefunction of the complex may be expanded
 \cite{DUBERNET:1991}
\begin{equation}
\Psi^{JM}_n=R^{-1}\sum_{jP\lambda\sigma}
\Phi^{JM}_{jP;\lambda\sigma}\,\chi^{J}_{Pn;j;\lambda\sigma}(R),
\label{fullwave} \end{equation} where the channel basis functions
are
\begin{eqnarray}
\Phi^{JM}_{jP;\lambda\sigma} = \varphi_\sigma \varphi_\lambda&&
\left({2j+1\over4\pi}\right)^\frac{1}{2}
D^{j*}_{P\omega}(\phi,\theta,0) \nonumber\\
\times&&\left({2J+1\over4\pi}\right)^\frac{1}{2}
D^{J*}_{MP}(\alpha,\beta,0), \label{chanbas}
\end{eqnarray}
The monomer basis functions are labelled by Hund's case (a)
quantum numbers $\lambda$ and $\sigma$, the projections of the
electronic orbital and spin angular momentum along the SH axis,
and $\omega=\lambda+\sigma$. \footnote{We adopt the convention of
using {\it lower-case} letters for all quantities that refer to
\emph{monomers}, and reserve upper-case letters for quantities
that refer to the complex as a whole.} $\lambda$, $\sigma$ and
$\omega$ are all signed quantities. The $D$ functions are Wigner
rotation matrices.\cite{Brink} The first $D$ function describes
the rotation of the monomer with respect to body-fixed axes, with
angular momentum quantum number $j$ (including electronic orbital
and spin angular momentum) and projection $P$ along the
intermolecular vector $\mathbf{R}$. The second $D$ function
describes the rotation of the complex as a whole, with total
angular momentum $J$ and projections $M$ and $P$ onto space-fixed
and body-fixed axes respectively. The angles ($\beta,\alpha)$
describe the orientation of the $\mathbf{R}$ vector in space.

The monomer Hamiltonian used here for SH($X^2\Pi$) is \cite{Brown}
\begin{equation}
H_{\rm mon} = b(\hat j-\hat l-\hat s)^2 + H_{\rm so},
\end{equation}
where the rotational constant $b$ is 9.465 cm$^{-1}$.
\cite{herzbergIV} For simplicity the spin-orbit Hamiltonian $H_{\rm
so}$ is taken to be independent of $R$ and $\theta$ and equal to
$a\lambda\sigma$, with $a=-378.5$ cm$^{-1}$.\cite{herzbergIV}

It is convenient to expand the sum and difference potentials in
terms of renormalized spherical harmonics $C_{lm}(\theta,\phi)$,
\begin{eqnarray}
V_0(R,\theta) = \sum_l V_{l0}(R) C_{l0}(\theta,0); \\
V_2(R,\theta) = \sum_l V_{l2}(R) C_{l2}(\theta,0).
\end{eqnarray}
The potential matrix elements between the angular basis functions
may then be written
\begin{eqnarray}&&\langle JM;jP;\lambda \sigma |\hat V|
JM;j'P';\lambda'\sigma'\rangle\nonumber\\
&&= \delta_{PP'} \delta_{\sigma\sigma'} \sum_l
V_{l,|\lambda-\lambda'|}(R) \, g_{l,\lambda-\lambda'}(j\omega;
j'\omega';P),
\end{eqnarray}
where the potential coupling coefficients are
\begin{eqnarray}
&&g_{l,\lambda-\lambda'}(j\omega; j'\omega';P)
\nonumber\\
&&\quad=(-1)^{P-\omega}\left[(2j+1)(2j'+1)
\right]^\frac{1}{2}\nonumber\\
&&\quad\times\left(\matrix{j&l&j'\cr
-\omega&\lambda-\lambda'&\omega'\cr}\right)
\left(\matrix{j&l&j'\cr -P&0&P\cr}\right).
\end{eqnarray}
The potential matrix elements are independent of $J$ and diagonal
in $P$. Nevertheless, in a full treatment the wavefunctions of the
complex are linear combinations of functions with different values
of $P$, because the operator $(\hat J - \hat\jmath)^2$ in Eq.\
\ref{ham} has matrix elements off-diagonal in $P$ ($\Delta
P=\pm1$). However, the full wavefunctions are eigenfunctions of
the parity operator: symmetrized basis functions may be
constructed by taking even and odd linear combinations of
$\Phi^{JM}_{jP;\lambda\sigma}$ and
$\Phi^{JM}_{j{-P};{-\lambda}{-\sigma}}$.

In the present work, the coupled equations are solved using the
BOUND program of Hutson.\cite{Hutson:bound:2006} The wavefunction
log-derivative matrix is propagated outwards from a boundary point
at short range ($R_{\rm min}$) and inwards from a boundary point
at long range ($R_{\rm max}$) to a matching point ($R_{\rm mid}$)
in the classically allowed region. If $E$ is an eigenvalue of the
Hamiltonian, the determinant of the difference between the two
log-derivative matrices at $R_{\rm mid}$ is zero.
\cite{Johnson:1978,Hutson:1994} The BOUND program locates
eigenvalues by searching for zeroes of the lowest eigenvalue of
the matching determinant,\cite{Hutson:1994} using bisection
followed by the secant method. In the present work we use $R_{\rm
min}=3.0$~\AA, $R_{\rm max}=9.5$~\AA, $R_{\rm mid}=4.2$~\AA\ and a
log-derivative sector size of 0.02 \AA. The basis set includes all
SH functions up to $j_{\rm max}=15/2$ in both spin-orbit
manifolds.

The energies obtained from full close-coupling calculations for
the lowest few $J=3/2$ levels of Ar--SH (actually carried out in
the equivalent space-fixed basis set \cite{DUBERNET:1991}) are
shown in Table \ref{tab:energies}. These levels all correlate with
SH $^2\Pi_{3/2}$, $j=3/2$ and are labelled with the projection
quantum number $P$ and Van der Waals stretching quantum number
$n$. We use the convention that levels in which $P$ and $\omega$
for the dominant basis functions have the \emph{same} sign are
labelled with positive $P$ and those where they have
\emph{different} sign are labelled with negative $P$.
\cite{DUBERNET:1991} In order of increasing energy, the lowest
four levels for Ar--SH have $P$ = +3/2, +1/2, $-3/2$, $-1/2$, in
contrast to Ar--OH where the order is +3/2, +1/2, $-1/2$, $-3/2$.
\cite{DUBERNET:1991, Dubernet:1993} The difference is due to the
anisotropy of the sum potential $V_0(R,\theta)$: the ratio
$V_{20}/V_{10}$ is larger for Ar--SH.

\begin{table}
\caption{Bound-state energies for $J=3/2$ levels of Ar--SH from
full close-coupling calculations (average $E_{\rm CC}$ and parity
splitting $\Delta E_{\rm CC}$), helicity decoupling calculations
($E_{\rm HD}$) and single-surface calculations on the lower
adiabatic surface including spin-orbit coupling ($E_{\rm Ad}$).
All energies are relative to the dissociation energy to form SH
($X^2\Pi_{3/2},\ j=3/2$). All energies are given as wavenumbers in
cm$^{-1}$. }\label{tab:energies}
\begin{ruledtabular}
\begin{tabular}{cccccc}
$P$ & $n$ & $E_{\rm CC}$ & $\Delta E_{\rm CC}$ & $E_{\rm HD}$ & $E_{\rm Ad}$ \\
$+3/2$ & 0 & $-102.745$ & $+3.5\times10^{-5}$ & $-102.652$ & $-102.725$  \\
$+1/2$ & 0 &  $-97.766$ & +0.144              &  $-97.593$ &  $-97.667$  \\
$-3/2$ & 0 &  $-94.940$ & $-1.1\times10^{-3}$ &  $-94.894$ &  $-95.035$  \\
$-1/2$ & 0 &  $-92.116$ & $-0.138$            &  $-92.222$ &  $-92.293$  \\
\\
$+3/2$ & 1 &  $-77.292$ & $+2.6\times10^{-5}$ &  $-77.111$ &  $-77.258$   \\
$+1/2$ & 1 &  $-72.276$ & +0.134              &  $-72.100$ &  $-72.148$   \\
$-3/2$ & 1 &  $-69.356$ & $-1.3\times10^{-3}$ &  $-69.265$ &  $-69.382$   \\
$-1/2$ & 1 &  $-67.207$ & $-0.124$            &  $-67.291$ &  $-67.315$   \\
\end{tabular}
\end{ruledtabular}
\end{table}

The close-coupling results may be compared with the microwave
experiments of Sumiyoshi \emph{et al.}\ ,\cite{sumiyoshi2000} who
obtained a rotational constant $B^{\rm eff}=1569.66$ MHz (0.05236
cm$^{-1}$) and parity doubling constant $q_J=0.32873$ MHz
($1.10\times10^{-5}$ cm$^{-1}$) for the ground state ($P=+3/2$).
These correspond to a $J=3/2-5/2$ separation of 0.262 cm$^{-1}$
and a $J=3/2$ parity splitting of $6.6\times10^{-5}$ cm$^{-1}$,
which compare with calculated values of 0.263 cm$^{-1}$ and
$3.5\times10^{-5}$ cm$^{-1}$ respectively. The very good agreement
for the rotational spacing suggests that the equilibrium distance
of our \emph{ab initio} potential is quite accurate. The
difference of almost a factor of 2 in the parity splitting is less
satisfactory, but Dubernet \emph{et al.}\ \cite{Dubernet:1992} have
shown that such terms involve complicated combinations of
high-order terms involving the difference potential, spin
uncoupling and Coriolis perturbations. Small differences between
the energies of excited states can have a large effect on the
parity splitting. Sumiyoshi \emph{et al.}\ \cite{sumiyoshi2005i}
have very recently measured microwave--millimetre-wave
double-resonance spectra of the $P=+1/2\leftarrow +3/2$ band of
Ar--SH. The centre of gravity of the parity components of the
$J=3/2\leftarrow 1/2$ line is 81.8 GHz (2.73 cm$^{-1}$). The
corresponding calculated quantity from our potential is 4.805
cm$^{-1}$. In addition, the measured parity splitting for the
$J=3/2$, $P=+1/2$ level is about 5300 MHz (0.177 cm$^{-1}$), which
compares with 0.144 cm$^{-1}$ from our calculations. An
interesting possibility for future work would be to adjust the
\emph{ab initio} potential to improve the fit to the spectroscopic
parameters using the morphing procedure of Meuwly and Hutson.
\cite{Meuwly:1999}

\subsection{Wavefunctions}

The full wavefunctions (Eq.\ \ref{fullwave}) contain contributions
from all possible values of $P$ and $\omega$ and are not separable
between the body-fixed angles ($\theta,\phi$) and the space-fixed
angles ($\beta,\alpha$). This makes them hard to visualize. In
addition, since the mixings depend on the total angular momentum
$J$, they are not convenient for calculating band intensities. We
therefore introduce two approximations to simplify the description
of the wavefunctions for this purpose. First, we introduce the
\emph{helicity decoupling} approximation, where matrix elements of
$(\hat J - \hat\jmath)^2$ off-diagonal in $P$ are neglected.
Secondly, we neglect matrix elements of $H_{\rm mon}$ off-diagonal
in $\sigma$ (spin-uncoupling terms). The coupled equations then
simplify to
\begin{widetext}
\begin{eqnarray}
\left[-\frac{\hbar^2}{2\mu} \frac{d^2}{dR^2} + E_{\omega j}^{\rm
mon} \right. &+& \left. \frac{\hbar^2}{2\mu
R^2}\left(J(J+1)+j(j+1)-2\omega^2\right)-E^J_{Pn}\right]
\chi^{J}_{Pn;j;\lambda\sigma}(R) \nonumber\\ &=& -\sum_{j'\lambda'}
\left<JM;j'P\lambda' \sigma\right|\hat V\left|JM;jP\lambda
\sigma\right> \chi^{J}_{Pn;j';\lambda'\sigma}(R). \label{eqcoup}
\end{eqnarray}
\end{widetext}
Since all matrix elements off-diagonal in $P$ and $\sigma$ have
been neglected, states with quantum numbers $(P,\omega)$ and
$(-P,-\omega$) are uncoupled and it is not necessary to take
combinations of definite total parity. However, states with
$(P,\omega)$ and $(-P,\omega)$ or $(P,-\omega)$ have different
potential energies and are nondegenerate.

The energy levels obtained from helicity decoupling calculations
for Ar--SH are included in Table \ref{tab:energies}. The
approximation is accurate to about 0.2 cm$^{-1}$ for $n=0$ and 1,
but is less reliable for higher states. In particular, the region
between $-60$ and $-40$ cm$^{-1}$ contains both $j=3/2,\ n=2$ and
$j=5/2,\ n=0$ levels. In the presence of the resulting
near-degeneracies, the terms that are omitted in the approximate
Hamiltonian can cause quite significant level shifts.

In the helicity decoupling approximation, $P$ is a good quantum
number. However, $\omega$ is not because $V_2$ mixes levels with
$\Delta\lambda=\pm2$ (but $\Delta\sigma=0$) and thus mixes
$\omega=+3/2$ with $-1/2$ and $\omega=+1/2$ with $-3/2$. However,
in the absence of terms off-diagonal in $\sigma$ the two sets are
not mixed with one another. Each wavefunction thus has only
\emph{two} components corresponding to different values of
$\omega$. The wavefunctions may be written
\begin{equation} \Psi^J_{Pn} = \sum_\omega
\chi^J_{Pn;\omega}(R,\theta) \, \Phi^{JM}_{P;\lambda\sigma},
\label{eqpsip}
\end{equation} where the basis functions now exclude the
$\theta$-dependence,
\begin{eqnarray}
\Phi^{JM}_{P;\lambda\sigma} = \varphi_\sigma \varphi_\lambda
\left(\frac{2J+1}{8\pi^2}\right)^{1/2}
D^{J*}_{MP}(\alpha,\beta,\phi),
\end{eqnarray}
and the 2-dimensional functions that characterize the components
of the wavefunction for each $\omega$ are
\begin{equation}
\chi^J_{Pn;\omega}(R,\theta) = \sum_j
\left(j+\textstyle\frac{1}{2}\right)^{1/2} d^j_{P\omega}(\theta)
\chi^{J}_{Pn;j;\lambda\sigma}(R), \label{eqchi}
\end{equation}
where $d^j_{j\omega}(\theta)$ is a reduced rotation matrix.
\cite{Brink}

\begin{figure*}
\includegraphics[width=\textwidth]{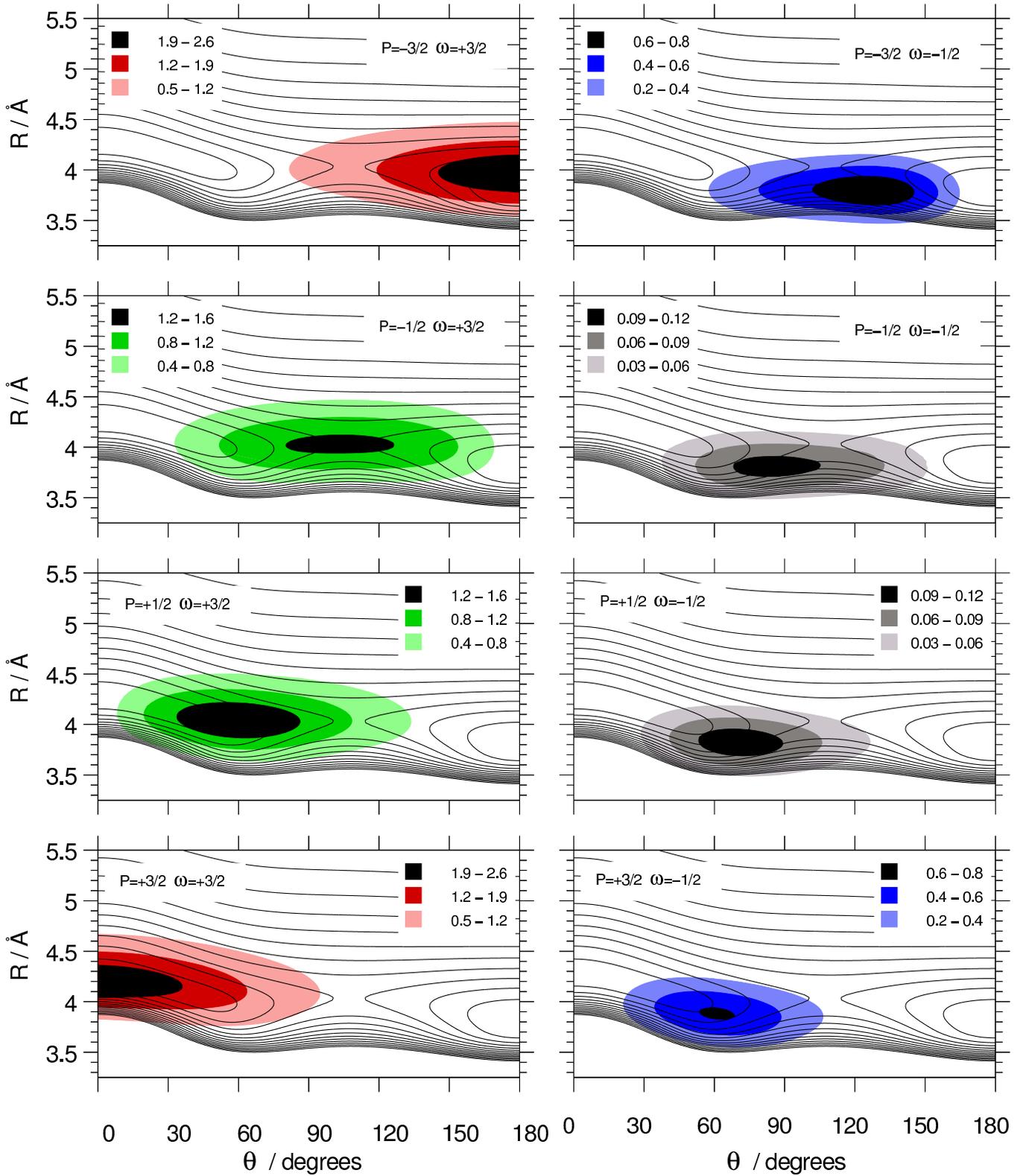}
\caption{[Colour Figure] Contour plots of the wavefunction
components, superimposed on the average potential.\label{wfnbig}}
\end{figure*}

We have adapted the BOUND program \cite{Hutson:bound:2006} to
calculate wavefunctions for this case by back-substituting into
the log-derivative propagation equations, as described for the
closed-shell (single surface) case by Thornley and Hutson.
\cite{Thornley:1994} Examples of the resulting wavefunctions
$\chi^J_{Pn;\omega}(R,\theta)$ are shown in Fig.\ \ref{wfnbig}. It
may be seen that the components for different values of $\omega$
have quite different radial and angular distributions.

The Ar--SH wavefunctions are qualitatively similar to those for
Ne--SH obtained by Cybulski {\em et al.}\ \cite{Cybulski:2000}
Since the potential anisotropy for Ar--SH is only a few tens of
cm$^{-1}$ in the well region, there is only weak mixing of SH
rotational functions with different values of $j$. For this reason
the wavefunctions are dominated by the functions
$d^{3/2}_{P\omega}(\theta)$, as described by Dubernet and Hutson
\cite{DUBERNET:1991} for the case of Ar--OH. The $d$ functions are
shown, for example, in Fig.\ 7 of ref.\ \onlinecite{DUBERNET:1991}
and the angular parts of the wavefunctions of Fig.\ \ref{wfnbig}
follow them quite closely.

\subsection{Adiabatic approximations}

\begin{figure}
\includegraphics[width=85mm]{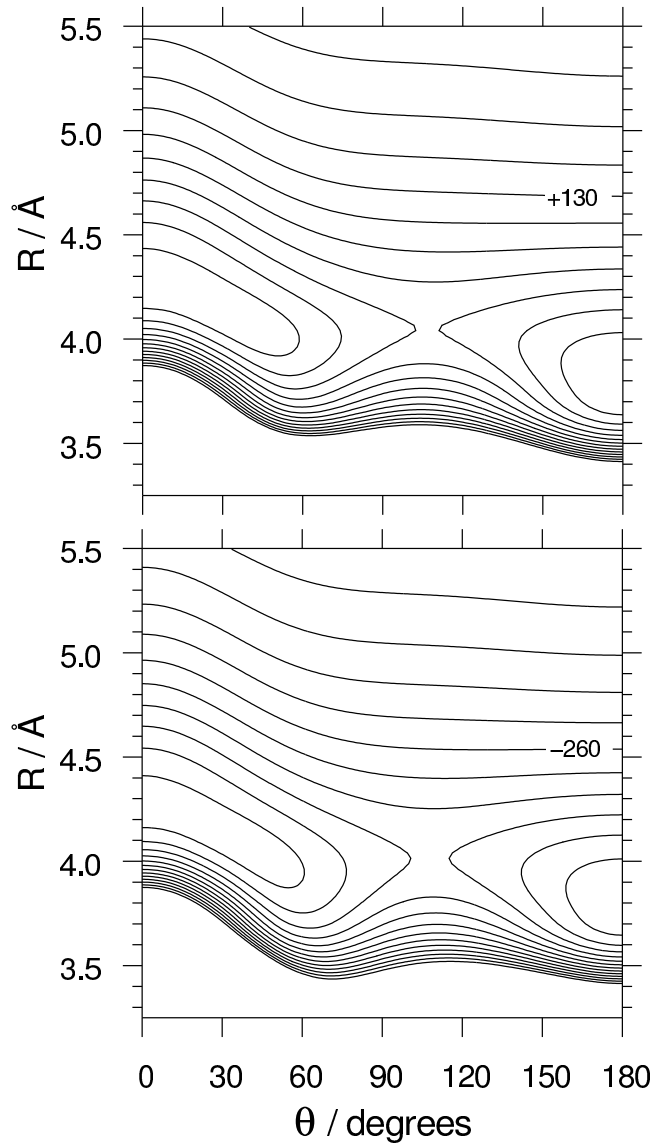}
\caption{Contour plots for Ar--SH ({\em X}) adiabats including
spin-orbit coupling. Contour lines are shown at 10 cm$^{-1}$
intervals, ranging from 70 cm$^{-1}$ to $+a/2$ for the upper
surface and from $-310$ cm$^{-1}$ to $-a/2$ for the lower surface.
\label{fig:so_adiabats}}
\end{figure}

\begin{figure}
\includegraphics[width=85mm]{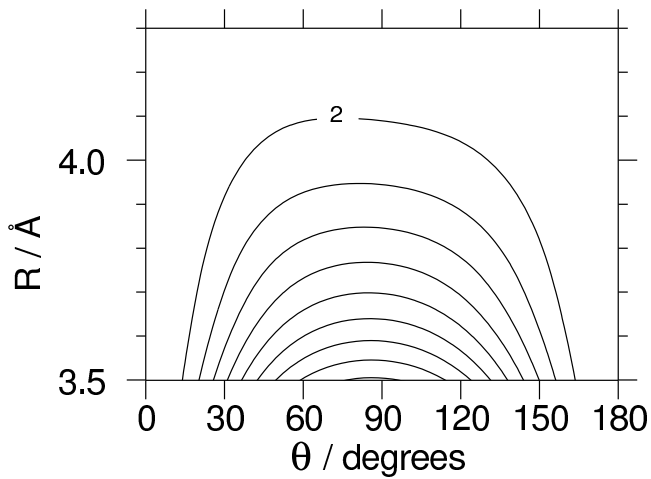}
\caption{Contour plot of the adiabatic mixing angle
$\alpha_{\textrm{ad}}$. This angle is derived from the adiabats
and is defined in Eq.\ \ref{anglead}. The contour lines are spaced
at $2^\circ$ intervals.\label{mix_ang_surf}}
\end{figure}

In a basis set of Hund's case (a) functions with signed values of
$\lambda=\pm1$ and $\sigma=\pm1/2$, we can define new adiabatic
surfaces (adiabats) including spin-orbit coupling as eigenvalues
of the Hamiltonian matrix at each value of $R$ and $\theta$,
\begin{equation} \left(\matrix{V_0+\frac{1}{2}a & 0 & V_2 & 0 \cr
              0 & V_0-\frac{1}{2}a & 0 & V_2 \cr
              V_2 & 0 & V_0-\frac{1}{2}a & 0 \cr
              0 & V_2 & 0 & V_0+\frac{1}{2}a \cr}\right),
\end{equation}
where again the spin-orbit Hamiltonian is taken to be simply
$a\lambda\sigma$. This clearly factorizes into two equivalent
$2\times2$ matrices, one containing $\omega=+3/2$ and $-1/2$ and
the other containing $\omega=+1/2$ and $-3/2$. The resulting
adiabats may be designated $V_+(R,\theta)$ and $V_-(R,\theta)$
with corresponding electronic functions $\psi_+(R,\theta)$ and
$\psi_-(R,\theta)$ given by
\begin{eqnarray} &&\left(\matrix{\psi_+(R,\theta) \cr \psi_-(R,\theta) \cr}\right)
\nonumber\\&&\quad=
\left(\matrix{\cos\alpha_{\rm ad}(R,\theta) & \sin\alpha_{\rm
ad}(R,\theta) \cr -\sin\alpha_{\rm ad}(R,\theta) & \cos\alpha_{\rm
ad}(R,\theta)}\right) \left(\matrix{\varphi_{\pm3/2} \cr
\varphi_{\mp1/2} \cr}\right), \label{anglead}
\end{eqnarray}
where $\varphi_\omega=\varphi_\lambda\varphi_\sigma$. The adiabats
for Ar--SH are shown in Fig.\ \ref{fig:so_adiabats}. Since for
Ar--SH $V_2(R,\theta)$ is small compared to $a$ in the well
region, the lower adiabat is always predominantly $\omega=\pm3/2$
in character and the upper adiabat is always predominantly
$\omega=\mp1/2$ in character. The corresponding mixing angle
$\alpha_{\rm ad}$ is 0 at $\theta=0$ and $180^\circ$ (where
$V_2(R,\theta)=0$) and less than $20^\circ$ at other angles for
$R>3.5$~\AA. A contour plot of the mixing angle is shown in Fig.\
\ref{mix_ang_surf}.

A further consequence of the large spin-orbit coupling constant is
that both adiabats resemble the {\it sum} potential
$V_0(R,\theta)$ much more than the $A'$ and $A''$ potentials. The
spin-orbit coupling has in effect quenched the splitting between
the $A'$ and $A"$ states. This explains why there is no tendency
for the wavefunctions shown in Fig.\ \ref{wfnbig} to ``fall into"
the non-linear minimum of the $A'$ state.

The existence of adiabats including spin-orbit coupling suggests a
Born-Oppenheimer separation in which the total wavefunctions are
written approximately as
\begin{equation} \Psi_{iPn} \approx R^{-1} \psi_i(R,\theta)
\chi_{iPn}(R,\theta), \label{psibo} \end{equation} where $\psi_i$
is one of the functions of Eq.\ \ref{anglead} and
$\chi_{in}(R,\theta)$ is a solution of an effective Schr\"odinger
equation of the form
\begin{eqnarray} &&\left[ -\frac{\hbar^2}{2\mu}\frac{\partial^2}{\partial R^2} + H_{\rm
rot}+ \frac{\hbar^2(\hat J -\hat\jmath)^2}{2\mu R^2} \right.
\nonumber\\ && +\ \Biggl. V_i(R,\theta) - E_{iPn}
\Biggr]\chi_{iPn}(R,\theta) = 0. \label{eqad}
\end{eqnarray} However, the appropriate angular operator $H_{\rm rot}$
to use in such a calculation is hard to define. The reduced rotation
matrices $d^j_{P\omega}(\theta)$ that describe the free SH molecule
are eigenfunctions of $H_{\rm rot} = b(\hat\jmath^2-2\omega^2)$,
where
\begin{equation}
\hat\jmath^2 = \left[-\frac{1}{\sin\theta}
\frac{\partial}{\partial\theta} \left(\sin\theta
\frac{\partial}{\partial\theta} \right) + \frac{P^2 + \omega^2
-2P\omega\cos\theta}{\sin^2\theta} \right]. \label{eqhrot}
\end{equation}
This contains singularities at $\theta=0$ and/or $180^\circ$ that
depend on the values of $P$ and $\omega$. However, there is no
single value of $\omega$ that is appropriate at all
configurations. The simplest approach is to replace $\omega$ in
Eq.\ \ref{eqhrot} with the value that is appropriate at $\theta=0$
and $180^\circ$ and solve Eq.\ \ref{eqad} in a basis set of $d$
functions for each value of $P$. This is equivalent to solving the
coupled equations using a basis set containing only functions with
a single value of $\omega$. The results obtained with this
approximation are included in Table \ref{tab:energies}: it may be
seen that it gives energies that are generally slightly too low
(compared to the helicity decoupling results), by 0.05 to 0.15
cm$^{-1}$. A slightly better but significantly more complicated
approximation would be to replace $\omega$ with
$\langle\omega\rangle$ and $\omega^2$ with
$\langle\omega^2\rangle$ in Eq.\ \ref{eqhrot} to give an improved
effective potential.

One approach that is clearly {\it not} appropriate is to carry out
a bound-state calculation on a single adiabat $V_\pm(R,\theta)$
assuming that the SH molecule behaves as a closed-shell rigid
rotor. Such a calculation would give substantially incorrect
energies and wavefunctions.

\begin{figure}
\includegraphics[width=85mm]{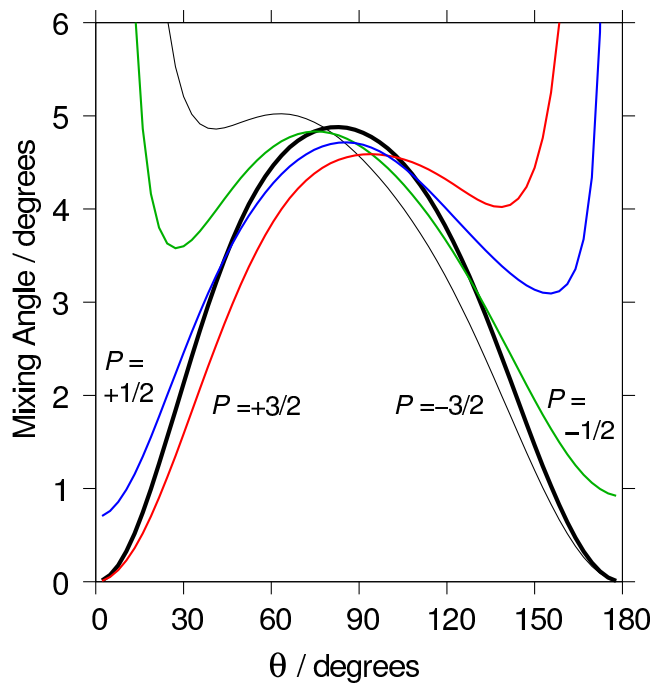}
\caption{[Colour Figure] Comparison of the adiabatic mixing angle
$\alpha_{\mathrm{ad}}$ (thick black line) with angles obtained
from wavefunctions correlating with $j=3/2$, $\omega=+3/2$ for
$P=+3/2$ (red), $P=+1/2$ (blue), $P=-1/2$ (green) and $P=-3/2$
(black). The mixing angles are shown as cuts through the
corresponding surfaces taken at $R=3.9$ \AA.
\label{mix_ang_wfns}}
\end{figure}

It is in fact true that {\it no wavefunction of the form
(\ref{psibo}) can have the correct behavior near both linear
geometries}. To see this, consider an alternative definition of
the mixing angle that can be obtained from a single wavefunction
in the helicity decoupling approximation,
\begin{equation} \tan\alpha^J_{Pn}(R,\theta) =
\frac{\chi^J_{Pn;\mp1/2}(R,\theta)}{\chi^J_{Pn;\pm3/2}(R,\theta)}.
\label{anglewav}
\end{equation}
This quantity is plotted for $n=0$ and all four $P$ values
corresponding to $j=3/2$ in Fig.\ \ref{mix_ang_wfns}. The mixing
angles for $P=+3/2$ and $P=+1/2$ bear some similarity to
$\alpha_{\rm ad}(R,\theta)$ (Fig.\ \ref{mix_ang_surf}) at small
$\theta$, but tend to $90^\circ$ instead of zero at $\theta =
180^\circ$. Conversely, the mixing angles for $P=-3/2$ and
$P=-1/2$ tend to $90^\circ$ at $\theta=0$. This is easy to explain
in terms of the reduced rotation matrices that appear in Eq.\
\ref{eqchi}. For example, the functions
$d^j_{\pm3/2,\pm3/2}(\theta)$ all behave as $\cos^3 (\theta/2)$ as
$\theta\rightarrow 180^\circ$, while the functions
$d^j_{\mp1/2,\pm3/2}(\theta)$ behave as $\cos (\theta/2)$. This
corresponds to $\tan\alpha^J_{3/2,n} \rightarrow \infty$ as
$\theta\rightarrow 180^\circ$ so $\alpha^J_{3/2,n} \rightarrow
90^\circ$ in that limit. The point here is that the component of
the $P=+3/2$ wavefunction on the $\omega=-1/2$ surface goes to
zero more slowly than that on the $\omega=+3/2$ surface as
$\theta\rightarrow 180^\circ$. Fig.\ \ref{mix_ang_surf} shows that
the coupled-channel wavefunctions (\ref{eqpsip}) for $P=+3/2$ and
$+1/2$ are predominantly in the $\omega=-1/2$ state near
$\theta=180^\circ$, which corresponds to the {\it upper} adiabat
rather than the lower one. The $P=-3/2$ and $-1/2$ wavefunctions
show similar behavior around $\theta=0$. This is not the behavior
implied by Eq.\ \ref{psibo}.

\section{Electronic spectrum calculation}

\subsection{Transition wavenumbers}

In order to calculate the line positions in the vibrationally
resolved $A\ ^2\Sigma^+ \leftarrow X\ ^2\Pi$ electronic spectrum,
we require the bound-state energies for the excited electronic
state, as well as those for the ground state. For the \emph{A}
$^2\Sigma^+$ state we make use of the recent PES presented by
Hirst \emph{et al.}\ \cite{hirst2004} This surface has a global
minimum of $-742.5$ cm$^{-1}$ at the linear Ar--H-S conformation
($\theta = 0^{\circ}$), and a secondary minimum of $-673.7$
cm$^{-1}$ for linear Ar--S-H ($\theta = 180^{\circ}$). The two
minima are separated by a barrier more than $600$ cm$^{-1}$ high
and the lowest-energy vibrational levels are localised within one
or other of the two wells.

The bound states of this PES have been analysed previously
\cite{hirst2004} and only a brief discussion is given here.
Bound-state energies were calculated as eigenvalues of the spin-free
triatomic Hamiltonian in Jacobi coordinates. Discrete variable
representations (DVRs) were employed for both the intermolecular
distance $R$ and the angle $\theta$. For $R$, 120 sinc-DVR functions
\cite{colbertmiller} were used, with DVR points ranging from 2.5
\AA\ to 8.5 \AA. For $\theta$, a 64-point DVR based on Legendre
polynomials was used. With this basis set, the bound levels of
interest were converged to at least seven significant figures. The
resulting levels are labeled by quantum numbers
($v_{\mathrm{SH}},b^K,n$), where $v_{\mathrm{SH}}$ and $n$ are
quantum numbers for the SH stretch and the atom--diatom stretch
respectively. $K$ is the projection of the total angular momentum of
the diatom, neglecting spin, onto the body-fixed $z$-axis, and $b$
is the number of nodes in the intermolecular angle $\theta$. The
resulting energies for levels with total angular momentum $N = 0$
(neglecting spin) are in precise agreement with previous results.
\cite{hirst2004} To facilitate the calculation of band intensities,
for $N = 1$ the helicity decoupled approximation was employed, in
which the Coriolis terms coupling different $K$-levels are ignored.
The helicity-decoupled energies are within 0.5 cm$^{-1}$ of the full
close-coupled results.\cite{hirst2004}

For the purpose of calculating transition frequencies, the
asymptotic separation of the potentials is taken to be the
experimental excitation energy from the $v=0,\ j=3/2$ level of the
$^2\Pi_{3/2}$ state to the lowest $v=0,\ j=1/2$ level of the
$^2\Sigma^{+}$ state of isolated SH, which is 30832.68 cm$^{-1}$.
\cite{ramsay52} All transitions of the complex were assumed to
originate from the $P=+3/2$ level of Ar--SH ($^2\Pi$). The
lowest-energy transition frequency for the complex is calculated
to be 30488.5 cm$^{-1}$, which is 31.5 cm$^{-1}$ greater than the
experimental value of 30457 cm$^{-1}$.\cite{miller97} This
agreement is reasonable, considering the level of theory used in
the calculation of the potentials.

\subsection{Transition dipole moments}
Calculations of spectroscopic intensities require transition
dipole moments $\mu_{\mathrm{tot}}^{\mathrm{if}}$, where
\begin{equation}
\mu_{\mathrm{tot}}^{\mathrm{if}} = \left< \mathrm{i}|
\mu_{\mathrm{el}} | \mathrm{f}\right>.
\end{equation}
The integrals involve the initial (i) and final (f) wavefunctions
as determined from bound-state calculations. In this work we
evaluated transition dipoles over \emph{internal} coordinates
($R,\theta$), neglecting overall rotation. This gives transition
dipoles that correspond to band intensities between intermolecular
vibrational states. The electronic dipole moment,
$\mu_{\mathrm{el}}$, is in general a parametric function of the
nuclear coordinates. In the body-fixed frame it may be expanded in
terms of reduced rotation matrices,
\begin{equation}
\label{eqn:elecdipmom} \mu_{\mathrm{el}}(R,\theta) =\sum_j
\mu_{\mathrm{el},j}^{\Delta\lambda}(R) d^j_{\Delta P, \Delta
\lambda}(\theta),
\end{equation}
where $\Delta P = P_{\mathrm{i}}-P_{\mathrm{f}}$ and $\Delta
\lambda = \lambda_{\mathrm{i}}-\lambda_{\mathrm{f}}$. In the
excited electronic state, $P=K\pm\textstyle{\frac{1}{2}}$.
In this work it is assumed that $\mu_{\mathrm{el}}$ consists
purely of contributions from the SH monomer, so that only $j=1$
contributes in Eq.\ \ref{eqn:elecdipmom} and the coefficients
$\mu_{\mathrm{el},j}^{\Delta\lambda}$ are independent of $R$.
Since we are dealing with a perpendicular transition in SH,
$\Delta\lambda=\pm1$. The transition dipoles were calculated as
one-dimensional Gaussian quadratures in $\theta$, then integrated
over $R$.

\subsection{Intensities and lifetime factors}

The signals in a pulsed-laser fluorescence excitation experiment
decay exponentially following each pulse, with a lifetime equal to
that of the excited state being probed. Intensities are typically
measured as the area under the decay curve, and so the
experimental intensities are proportional to the lifetime of the
excited state. Ar--SH is somewhat unusual in the large range of
lifetimes exhibited by different vibrational levels in the
\emph{A} state. It is known that the presence of the Ar atom
blocks the electronic predissociation of the SH radical, leading
to a greatly increased lifetime of up to 600~ns for low-lying
bound levels, compared to $\sim 1$ ns for the uncomplexed species.
\cite{mccoy98, miller97} However, the actual lifetime depends on
the degree of vibrational excitation, and lifetimes specific to
particular levels have been calculated by McCoy.\cite{mccoy98}

\begin{figure}
\includegraphics[width=85mm]{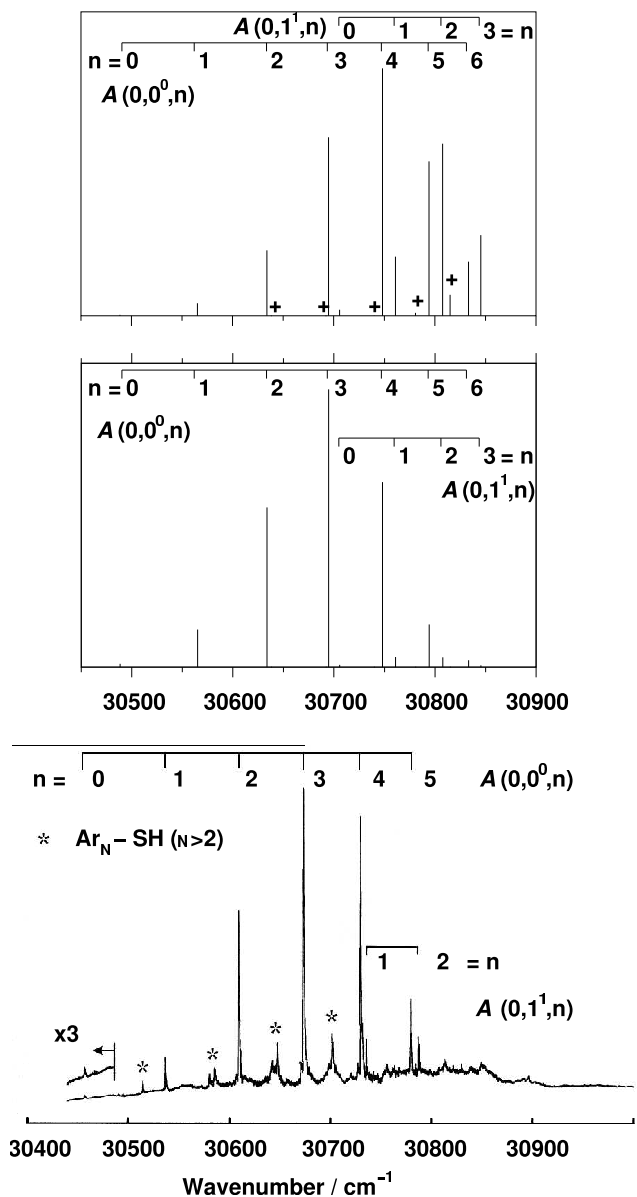}
\caption{Calculated vibrationally resolved fluorescence-excitation
spectrum of Ar--SH for the \emph{A} $^2\Sigma^+$ -- \emph{X}
$^2\Pi$ electronic transition. The lifetime weighting of the intensities
is absent for the the top panel, and present for the
middle panel. Lines labeled with $+$ indicate transitions to the
$\theta=180^{\circ}$ well of the \emph{A} state.
For comparison, the experimental spectrum is shown in the
bottom panel, taken from Ref.\ \onlinecite{miller97}.
Note that the experimental spectrum contains contributions from
Ar$_2$--SH and larger clusters as well as Ar--SH.\label{fig:simspec}}
\end{figure}

The top panel of Fig.\ \ref{fig:simspec} shows a spectrum
calculated directly from the squares of transition dipoles, while
the center panel shows a spectrum in which the intensities have
been multiplied by McCoy's lifetime values. Clearly this is
possible only for levels for which lifetime data exist, and
transitions to other levels are omitted in the center panel
(\emph{i.e.}, it is assumed that their lifetimes are small). Also
shown is an experimental spectrum from Ref.\
\onlinecite{miller97}. The spacings between the peaks in the
calculated Ar--SH spectrum are consistently $\sim 5$\% smaller
than in experiment. It is clear that the intensity distributions
are significantly different in the two calculated spectra, and
that the one that includes lifetime factors gives considerably
better agreement with experiment. The agreement in intensities is
quite good, especially considering that the experimental spectrum
was most likely not normalised for dye laser power.
\cite{millerperscomm} From our results it seems likely that the
small peak at $\sim 30810$ cm$^{-1}$ in the experimental spectrum
can be assigned to the transition to $(0,0^0,6)$.

Even without the lifetime weighting, transitions to levels localized
in the $\theta=180^{\circ}$ well of the \emph{A} state, which are
labelled with $+$ symbols in Fig.\ \ref{fig:simspec}, are weak in
the simulated spectrum. This arises because of poor overlap with the
$P=+3/2$ ground-state wavefunction (which is concentrated around
$\theta=0$). In addition, it is likely that such levels have short
lifetimes close to that of uncomplexed SH \cite{hirst2004} and so
will have even lower intensities in the fluorescence excitation
spectrum. These levels have not been observed experimentally to our
knowledge.

\section{Summary}

We have obtained new \emph{ab initio} potential energy surfaces
for the Ar--SH complex in its ground $^2\Pi$ electronic state and
used them to calculate bound-state energies and wavefunctions
using coupled-channel methods. We have also described a new
adiabatic approximation that includes spin-orbit coupling and can
be used to calculate bound states on a single potential energy
surface. However, the adiabatic wavefunctions fail to reproduce
some features of the true wavefunctions. We have used our results
to simulate the vibrationally resolved laser-induced fluorescence
excitation spectrum of Ar--SH, with intensities modelled using
calculated transition dipole moments and calculated lifetimes. The
inclusion of the lifetime factor is important to obtain
satisfactory agreement with the experimental intensities.

\section{Acknowledgements}
The authors would like to thank Terry Miller for providing his
experimental spectrum. RJD also thanks Dr. Stuart Mackenzie for
helpful discussions, and is grateful to the Engineering and
Physical Sciences Research Council (EPSRC) for funding.

\bibliography{arsh}

\end{document}